\documentclass{emulateapj}

\usepackage{apjfonts}
\usepackage{natbib}

\newcommand{\msun}{M_{\sun}}

\newcommand{\degree}{\,^\circ}

\catcode`\@=11
\newcommand{\gapprox}{\mathrel{\mathpalette\@versim>}}
\newcommand{\lapprox}{\mathrel{\mathpalette\@versim<}}
\newcommand{\propapprox}{\mathrel{\mathpalette\@versim\propto}}
\newcommand{\@versim}[2]
  {\lower3.1truept\vbox{\baselineskip0pt\lineskip0.5truept
\ialign{$\m@th#1\hfil##\hfil$\crcr#2\crcr\sim\crcr}}}
\catcode`\@=12

\shorttitle{COORDINATED X-RAY AND RADIO OBSERVATIONS OF RRAT J1819--1458}
\shortauthors{MILLER ET AL.}

\begin{document}

\title{Simultaneous X-ray and Radio Observations of Rotating Radio Transient J1819--1458}

\author{J.~J.~Miller\altaffilmark{1}, M.~A.~McLaughlin\altaffilmark{1,2},
N.~Rea\altaffilmark{3}, K.~Lazaridis\altaffilmark{4}, 
E.~F.~Keane\altaffilmark{4},
M.~Kramer\altaffilmark{4,5}, A.~Lyne\altaffilmark{5}}

\altaffiltext{1}{Department of Physics, West Virginia University, Morgantown, WV 26506.}
\altaffiltext{2}{Also adjunct at National Radio Astronomy Observatory, Green Bank, WV 24944.}
\altaffiltext{3}{Institut de Ci\`encies de l'Espai (IEEC-CSIC) Campus UAB, Fac. de Ci\`encies, Torre C5, parell, 2a planta 08193 Barcelona,  Spain}
\altaffiltext{4}{Max Planck Institut f\"{u}r Radioastronomie, Auf dem H\"{u}gel 69, 53121 Bonn, Germany}
\altaffiltext{5}{Jodrell Bank Centre for Astrophysics, School of Physics and Astronomy, University of Manchester, Manchester M13 9PL, UK}

\vskip 1 truein

\newpage

\begin{abstract}
We present the results of simultaneous radio and X-ray observations of PSR~J1819$-$1458.
Our 94-ks {\it XMM-Newton} observation of the high magnetic field ($\sim$5$\times10^{13}$~G) pulsar
reveals a blackbody spectrum (kT $\sim$130~eV) with 
a broad absorption feature, possibly composed of
two lines at $\sim$1.0 and $\sim$1.3~keV.
We performed a correlation analysis of the X-ray photons
with radio pulses detected in 
16.2 hours of simultaneous observations at $1-2$~GHz with
the Green Bank, Effelsberg, and Parkes telescopes, respectively.
Both the detected X-ray photons and radio pulses appear to be randomly distributed in time.
We find tentative evidence for a correlation between the detected
radio pulses and X-ray photons on timescales of less than 10 pulsar spin periods,
with the probability of this occurring by chance being 0.46\%.
This suggests that the physical process producing the radio pulses may also heat the polar-cap.
\end{abstract}


\section{Introduction}
\label{intro}
There are over 2000 known pulsars\footnote{http://www.atnf.csiro.au/people/pulsar/psrcat/},
with roughly 70 of these labeled as Rotating Radio Transients
(RRATs)\footnote{http://www.as.wvu.edu/$\sim$pulsar/rratalog};
see \citet{km11} for a recent review.
The single pulses of RRATs have similar widths and intensities to single pulses of other pulsars,
but despite an underlying periodicity at the neutron star's rotational period, radio pulses are sporadically detected.
It is unclear why the emission of these objects is so sporadic, and
numerous theories have been put forward which rely on both 
internal factors, such as RRATs may be dying or extreme nulling pulsars \citep{zgd07},
or external factors such as modulation of the emitted pulses from a radiation belt similar to planetary magnetospheres \citep{lm07}
or disturbances from the pulsar's asteroid belt \citep{cs08}.

PSR~J1819$-$1458 has a spin period of $P=4.26$~s, with a pulse detected roughly
every three minutes in Parkes observations 
above a flux density of $S=0.6$~mJy at 1.4~GHz \citep{mll+06}.
It has a characteristic age of $\tau_c=117$~kyr,
a spin-down luminosity of $\dot{E}_{\rm rot} = 3\times10^{32}$~ergs~s$^{-1}$,
and a high inferred surface magnetic field strength of 
$B=5\times10^{13}$~G at the magnetic equator (see, e.g., \citet{lk05} for a definition of these terms).
The distance estimate from its dispersion measure (DM) of 196.0$\pm0.4$~pc~cm$^{-3}$ \citep{ezy+08}
and the Galactic electron density model of \citet{cl02} is 3.6~kpc
with considerable (at least 25\%) uncertainty.

A previous 43-ks observation of PSR~J1819$-$1458 by {\it XMM-Newton} \citep{mrg+07}
found best-fit spectral models
with neutral hydrogen column densities
$N_{\rm H} \sim$7$ \times 10^{21}$ cm$^{-2}$, 
temperatures near $kT\sim0.14$~keV,
a single absorption line near $\sim$1~keV,
and unabsorbed fluxes $\sim$2$ \times 10^{-12}$~ergs~cm$^{-2}$~s$^{-1}$ ($0.3-5$~keV),
which yield a blackbody emission radius (at infinity assuming a 3.6~kpc distance)
of $R=8^{+7}_{-3}$~km.
This temperature is expected for a $10^4-10^5$~year-old cooling neutron star's emission \citep{yp04},
generally in agreement with PSR~J1819$-$1458's age.
The unabsorbed fluxes yield luminosities which exceed
the spin-down luminosity of PSR~J1819$-$1458 by a factor of
$L_{\rm 0.3-5.0 keV}/\dot{E}_{\rm rot} = \sim$4$ \times10^{33}/3\times10^{32} \simeq 6-18$,
depending on the spectral model,
which is possible given the thermal origin of the X-ray emission.
The results reported by \citet{mrg+07} are consistent with
both a chance {\it Chandra} observation of PSR~J1819$-$1458 \citep{rbg+06}
as well as deeper {\it Chandra} observations of PSR~J1819$-$1458 \citep{rmg+09,crb+12}.
The absorption line seen with {\it XMM-Newton} by \citet{mrg+07} was confirmed with {\it Chandra} by \citet{rmg+09},
which rules out instrumentation as the cause.
The latter Chandra observations also
revealed a bright pulsar wind nebula around PSR~J1819$-$1458,
with an inferred X-ray efficiency of
$\eta_{X} = L_{\rm pwn:0.5-8.0 keV}/\dot{E}_{\rm rot} = 6.0\times10^{31}/3\times10^{32} \simeq 0.2$.

Several spectral models can be used to explain the absorption in the spectrum of PSR~J1819$-$1458.
Possible explanations are elements in the interstellar medium (ISM),
elements in the neutron star's atmosphere, or cyclotron absorption.
A cyclotron proton resonant scattering model yields the magnetic field strength
$B_{\rm cy}=1.6E_{\rm cy}({\rm keV})/y_{\rm G}~10^{14}$~G,
where $E_{\rm cy}$ is the cyclotron proton energy,
and $y_{\rm G} = (1-2GM/c^2R)^{1/2}$ is the gravitational redshift factor
($\sim$0.77 using a canonical neutron star mass $M = 1.4 \msun$, and neutron star radius $R = 10$~km).
If the cyclotron resonance was due to electrons and not protons,
the inferred magnetic field strength would be $m_p/m_e = 1.8\times10^{3}$ times weaker,
making proton cyclotron resonance more likely due to the high inferred
surface magnetic field strength of PSR~J1819$-$1458.

Absorption lines have been observed in several other isolated neutron stars.
These include some X-ray Isolated Neutron Stars
(\citet{hhv+12}; see \citet{t09} and \citet{kk11} for recent reviews),
which have absorption lines reported at lower energies (300$-$700~eV) than those observed for PSR~J1819$-$1458.
Furthermore, another rotation-driven pulsar, PSR~J1740+1000,
has been shown to have an absorption feature around 600~eV \citep{kdmp12}.
It is unclear why some neutron stars exhibit absorption and others 
do not, with various explanations offered for these absorption lines,
e.g. proton cyclotron resonances and atomic transitions in light elements \citep{t09,kk11}.

While spectral observations are important for probing
the pulsar environment, X-ray timing observations can
also be useful to learn about emission mechanisms,
especially when combined with synchronous radio observations.
If radio pulses are correlated with X-ray photons, 
then a combined mechanism could be responsible for
radio and high-energy emission. Such tests have been done to correlate
radio giant pulses from the Crab pulsar pulses with non-thermal X-ray and gamma-ray photons. 
No correlation was found in these studies \citep{bkm+11} but a
correlation was found between radio giant pulses and optical emission \citep{sso+03}, 
suggesting an overall increase in particle density could be
responsible for the giant pulses. 
In the case of RRATs, \citet{zgd07} suggest that 
we should expect an increase in both non-thermal and thermal X-ray emission
close to radio pulse detection times if their
sporadicity is due to their reactivation model.
This model suggests that the pulsar is only active
when the conditions in its magnetosphere allow for pair production 
which instigates coherent radio emission 
that results in non-thermal X-ray photons and thermal emission from polar-cap heating.
If the pulsar is always active and the sporadicity is due to radio emission direction reversal, however,
then the non-thermal and polar-cap heating will always be present
and we should therefore 
not see an increase in X-ray emission close to radio pulse detection times.

We were awarded 94~ks of {\it XMM-Newton} time to improve the accuracy of the
spectral parameters, determine the origin of the absorption lines,
search for evidence of a non-thermal power-law structure in the spectrum,
and explore whether the X-ray and radio emission is correlated. 
We were also awarded time on the Robert C. Byrd Green Bank Telescope (GBT), 
the Parkes radio telescope, and the Effelsberg radio telescope
for simultaneous radio observations.
We report here on the results of these observations.
In Section~\ref{xrayprops} we describe the X-ray properties of PSR~J1819$-$1458,
quantifying absorption features and the possibility of a power-law tail.
In Section~\ref{radio} we describe the star's radio properties.
In Section~\ref{correlate} we compare the observed profile at both wavelengths
and present the correlation of pulse arrival times.
Finally, we draw some conclusions in Section~\ref{conclusions}.

\section{X-Ray Observations and Analysis}
\label{xrayprops}

We observed PSR~J1819$-$1458 with {\it XMM-Newton} for 94~ks on 31 March 2008.
These data were taken with EPIC-PN in Full Frame mode and the two MOS with the central CCD in Small Window mode,
as was done by \citet{mrg+07}.
The time resolutions of the EPIC-PN in Full Frame mode and two MOS CCDs in Small Window mode
are 73.4~ms and 0.3~s, respectively.
PSR~J1819$-$1458 appeared as a point source with the following J2000 coordinates:
right ascension $\alpha = 18^{\rm h}19^{\rm m} 34^{\rm s}$
and declination $\delta = -14\degree 58\arcmin 04\arcsec$
($4\arcsec$ error in each coordinate, where all the errors in this paper are stated at the 1$\sigma$ confidence level),
consistent with previous X-ray observations
and the position derived from radio timing.
We could not distinguish the $\sim$5.5$\arcsec$ extended emission region
detected by \citet{rmg+09} with {\it Chandra},
which has a spatial resolution of $\sim$0.5$\arcsec$ \citep{wbc+02,gbf+03}
compared to the $\sim$6$\arcsec$ spatial resolution of {\it XMM-Newton} \citep{wsf+09}.

The timing and spectral analyses were done using the {\it XMM-Newton}
Scientific Analysis System (SAS) tools\footnote{http://xmm.esa.int/sas/}, version 12.7.0.
The Current Calibration File (CCF) was built using the {\tt cifbuild} command
on the SAS tools website\footnote{http://xmm.vilspa.esa.es/external/xmm\_sw\_cal/calib/cifbuild.shtml},
using the observation date 2008-03-31T14:06:38.
In order to exclude events not associated with the pulsar, e.g. solar flares,
we defined good time intervals (GTIs) by binning all of the PN and MOS detection times,
as well as the detection times from within a 20$\arcsec$ circular radius centered on the source position,
into ten-second intervals.
We then identified time intervals with excessive photon counts which were not confined to the source region,
as determined by visual inspection (areas dominated by non-zero baselines
and count rates greater than 100 counts per 10~second time intervals on the PN detector),
and excluded those time intervals from the GTIs and our further analysis.
Multiple GTI ranges were tested for both timing and spectral analysis.
Our analysis resulted in three GTIs for each of the PN, MOS1, and MOS2 detectors.
This excluded large bursts at the beginning and end of the observation which were not confined to the source regions;
these GTIs span 68.6~ks (19~hrs) from MJD~54556.8 to MJD~54557.6 with each detector having 
three small interruptions, each spanning 3.5$-$15.6~seconds, as shown in Table~\ref{GTItable}.
The GTI and photon arrival times were barycentered to the center of the solar system using the XMM analysis tool {\tt barycen}
and the X-ray derived position.

\subsection{Timing Analysis}
\label{timing}

Our time resolution is sufficient for studying the pulse profile
because of the long period of the pulsar.
For timing analysis we included all PN and MOS events within the GTIs satisfying 
a PATTERN $\le12$ requirement (i.e. allowing for single, double, triple, and quadruple events).
To ensure extraction of at least 90\% of the source photons,
we chose a $20\arcsec$ circular radius centered on the source position in the data.
We also extracted background counts from four nearby $20\arcsec$ circular region
free of point sources and on the same central CCD as the source region to
measure the average background rate.
The photon arrival times were folded with the radio timing ephemeris of PSR~J1819$-$1458 using
{TEMPO}\footnote{http://www.atnf.csiro.au/research/pulsar/tempo/}.
The data were folded for a combination of 99 trial values of the number of pulse phase bins (2$-$100 bins),
1969 values of minimum energy $E_{\rm min}$ (0.155$-$9.995~keV), and 1969 values of maximum energy $E_{\rm max}$ (0.160$-$10~keV),
creating $\sim384$~million profiles.
For each trial number of phase bins, value of $E_{\rm min}$, and value of $E_{\rm max}$,
a $\chi^2$ value was calculated for a fit of the folded profile to a flat (i.e. random) distribution.
The background rate subtracted X-ray profile with the lowest probability of being drawn from a flat 
distribution, $P=10^{-52.1}$, has ten phase bins,
$E_{\rm min}$ = 0.5~keV, and $E_{\rm max}$ = 2.6~keV and is shown in Figure~\ref{fig_profile}.
Of the 6630 total PN photons within the $20\arcsec$ radius, 5692 fall within this energy range.

A sinusoid was fit to the X-ray profile to determine the peak using a least-squares fitting routine.
When adding a second-order sinusoid to the fit,
$f(t) = A_{1}\cos(2\pi(x-\phi_{0})) + A_{2}\cos(4\pi(x-\phi_{0}))$,
the reduced $\chi^2$ is decreased from 2.3 to 1.0.
Similarly, fitting a Gaussian function produces a fit with reduced $\chi^2 = 1.0$.
The phase of the peaks of both the double sinusoid and the Gaussian function was 
$0.02\pm0.01$, where phase zero is the peak of the radio pulse profile.
These fits are also shown as the dotted and dashed line in Figure~\ref{fig_profile}, respectively.

The X-ray pulse profile has a 0.5$-$2.6 keV intrinsic pulsed fraction, 
defined as $(F_{\rm max}-F_{\rm min})/(F_{\rm max}+F_{\rm min})$, 
where $F_{\rm max}$ and $F_{\rm min}$ are the minimum and maximum background-corrected counts of the X-ray pulse profile, 
of (33.9$\pm$0.9)\%, using ten bins and assuming Poisson (i.e. $\sqrt{N}$) errors.
Previous background-corrected pulsed fractions reported for PSR~J1819$-$1458 are
(34$\pm$6)\%, (28$\pm$7)\%, and (49$\pm$10)\% for the 0.3$-$5~keV, 0.3$-$1~keV, and 1$-$5~keV energy ranges, respectively \citep{mrg+07},
and (37$\pm$3)\% for the 0.3$-$5~keV energy range \citep{rmg+09}.
Pulsed fractions measured within the same energy ranges with our data set yields
(31.0$\pm$0.8)\%, (30$\pm$1)\%, and (49$\pm$1)\% for 0.3$-$5~keV, 0.3$-$1~keV, and 1$-$5~keV, respectively.

\subsection{Spectral Analysis}
\label{spectrum_analysis}

\begin{figure*}
  \includegraphics[angle=0,width=7.0in]{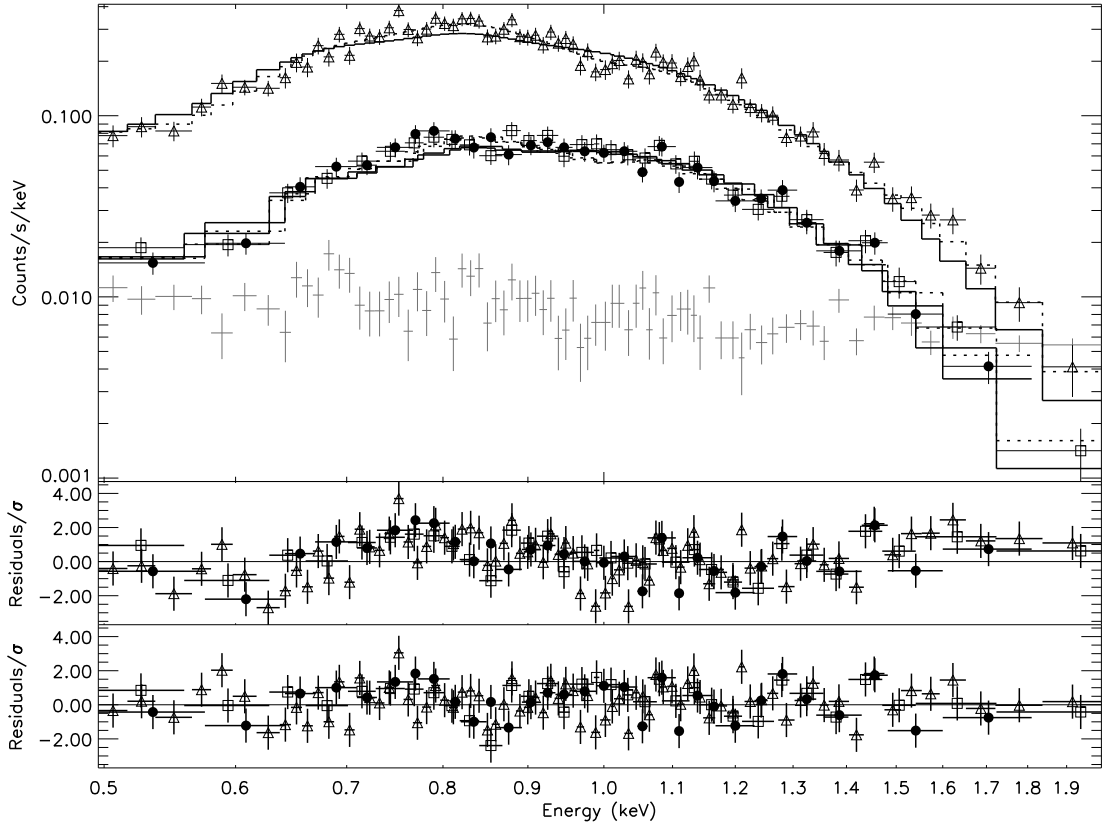}
  \caption{
  X-ray spectrum of PSR~J1819$-$1458
  using photons with energies in the 0.5$-$2.0~keV range
  and PATTERN~$\le4$
  from both our observation and \citet{mrg+07}.
  Data have been rebinned for plotting purposes by a factor of two 
  from 157, 63, and 66 bins to 78, 30, and 33 bins for PN, MOS1, and MOS2, respectively.
  {\it Top:} The dark crosshairs indicate the PN ({\it triangles}),
  MOS1 ({\it filled circles}) and MOS2 ({\it squares})
  source spectra, respectively.
  The light crosshairs represent the PN background spectrum.
  The solid lines indicate the simplest model fit,
  a blackbody with interstellar absorption, an underabundance of oxygen,
  and solar abundances from \citet{l03} for elements other than hydrogen and oxygen ({\tt vphabs*bbody});
  while the dotted lines indicate one of the best model fits ({\tt vphabs*gabs*gabs*bbody}),
  which also includes two Gaussian absorption lines around 1.0 and 1.3~keV.
  {\it Middle:} Normalized PN ({\it triangles}), MOS1 ({\it filled circles}) and MOS2 ({\it squares}) residuals for the {\tt vphabs*bbody} model.
  {\it Bottom:} Normalized PN ({\it triangles}), MOS1 ({\it filled circles}) and MOS2 ({\it squares}) residuals for the {\tt vphabs*gabs*gabs*bbody} model.
  }
  \label{spectrum_all}
\end{figure*}

For spectral analysis, we selected photons from the PN detectors with a more stringent 
${\rm PATTERN}\le4$ requirement (i.e. allowing for single and double events), 
as the background will affect results more significantly.
As in the timing analysis, 
we extracted the source photons from within a 20$\arcsec$ circular radius centered on the source position,
which yielded 6974 total events in a 0.5$-$2.0~keV energy range.
We also extracted background counts from four 20$\arcsec$ circular regions centered on off-source positions
free of point sources and on the same central CCD as the source region.
The spectrum was then rebinned so that there were at least 30 counts per spectral bin
so that we could use the $\chi^2$ statistic\footnote{https://heasarc.gsfc.nasa.gov/xanadu/xspec/manual/manual.html}.
Additionally, we similarly processed the data from \citet{mrg+07} and added the two observations together
with the XSPEC command {\tt mathpha} with a Gaussian error propagation method
to create the spectrum shown in Figure~\ref{spectrum_all}.
We also processed the MOS1 and MOS2 detections from the observation as well as from \citet{mrg+07}
also shown in Figure~\ref{spectrum_all}.
We will only discuss PN spectral analysis hereafter, 
but both of our MOS spectra model fits are in agreement with the PN spectral analysis.

We restricted the energy range of our spectral fitting to 0.5$-$2.0~keV.
This is narrower than that used for timing as at higher energies,
the spectrum count rates were comparable to the background region count rates.
We were unable to fit a spectral model with $\chi^{2}<2$
without addressing a feature in the residuals of the fits near $\sim$0.5~keV.
\citet{mrg+07} ignored the 0.5~keV feature by excluding the 0.50$-$0.53~keV energy range 
from their spectral fitting, but mentioned that an underabundance of oxygen could explain it.
We found that the oxygen edge in the XSPEC model {\tt vphabs} fit
our $\sim$0.5~keV feature well and included it as well as the 0.50$-$0.53~keV
energy range in all of our model fits.
Solar abundances from \citet{l03} were assumed for elements other than Neon and Oxygen.
We also investigated fitting our spectral models to energy ranges above 2.0~keV, 
where it looked like a possible power-law tail may have been present,
but attempts to fit the background dominated portion of the spectrum
yielded unacceptable $\chi^2$ values.

Modeling the blackbody spectrum without fitting for the 1.0~keV feature results in $\chi^2\sim$1.4.
Adding an absorption model around 1.0~keV,
modeled as either an empirical Gaussian absorption
or as cyclotron absorption,
yields a better fit, $\chi^2\sim$1.2 (see Table~\ref{spec}).
Using an underabundance of neon to explain this feature as was done by \citet{mrg+07} 
does not yield as good a fit as either the Gaussian or cyclotron absorption.
While it is possible that the spectrum could also consist of
two blackbody components, cooler emission from the surface along with a smaller hotspot,
fitting yielded $\chi^2=1.19$ {but with large blackbody emission radii errors}, (see Table~\ref{spec}).
Furthermore, the residuals suggest a second feature around 1.3~keV, 
so we tried to add another Gaussian absorption line to the model,
resulting in $\chi^2=1.09$.
We ran Monte Carlo simulations to assess the significance of the addition of the second absorption 
feature (see \citet{roz+05} for further details), 
and found a significance of $\sim$3$\sigma$ for its addition to the continuum plus one feature model,
i.e. $>99\%$ likelihood of two absorption lines rather than just one. 
The two-line model is then preferred at a 3$\sigma$ significance, with $\chi^2=1.09$ (see Table~\ref{spec}).
We did not include the two cyclotron absorption model
even though it fit equally well as the others because the two energies
are not harmonically related.
We also tested XSPEC neutron star atmosphere model {\tt nsa}
which yielded parameters in agreement
with those found in Table~\ref{spec}.
We performed a phase-resolved analysis, dividing the observation in the on-pulse spectra
(0.0$-$0.25 and 0.75$-$1.0 pulse phase in Figure~\ref{fig_profile})
and off-pulse spectra (0.25$-$0.75 pulse phase in Figure~\ref{fig_profile}).
Results of the spectral fits to the on- and off-pulse spectra agreed with the parameters fit
to the phase-integrated spectra within the parameter uncertainties.

The Leiden/Argentine/Bonn Survey of Galactic HI map \citep{kbh+05} and 
Dickey and Lockman HI in the Galaxy map \citep{dl90}
quote the total hydrogen column density along the line of sight of PSR~J1819$-$1458
as $1.25\times10^{22}$cm$^{-2}$ and $1.64\times10^{22}$cm$^{-2}$, respectively,
using a weighted average of all points within one degree of PSR~J1819$-$1458.
Since the maps represent column densities along the entire line of sight
including hydrogen beyond the pulsar, it is reassuring that the 
hydrogen column densities in Table~\ref{spec} are generally less than the map measurements.
\citet{hnk13} found an empirical relationship between $N_{\rm H}$ and DM for radio pulsars
of $N_{\rm H}$ (10$^{20}$~cm$^{-2}$) $= 0.30^{+0.13}_{-0.09}$DM (pc\,cm$^{-3}$),
which implies an average radio pulsar ionization rate of $10^{+4}_{-3}$\%.
When applied to J1819$-$1458, this relation implies 
$N_{\rm H} = 0.30^{+0.13}_{-0.09} \times 196.0\pm0.4$~pc~cm$^{-3} = 0.6^{+0.3}_{-0.2}\times10^{22}$cm$^{-2}$,
which agrees with three of the six fitted models
shown in Table~\ref{spec}.

\section{Radio Observations and Analysis}
\label{radio}

Radio observations were carried out contemporaneously with the {\it XMM-Newton} satellite observations.
The first radio observations were performed with the 64-m Parkes radio telescope
located in NSW, Australia using the multibeam receiver.
After PSR~J1819$-$1458 set at Parkes, we continued observing the source with
the 100-m Effelsberg radio telescope located in Effelsberg, Germany.
Just before the Effelsberg observations ended,
we started observing PSR~J1819$-$1458 with the 105-m GBT in Green Bank, WV, USA.
The GBT measurements were followed up once again with the Parkes radio telescope
for the remaining hour of the scheduled {\it XMM-Newton} observations.
The durations and parameters of each radio observation are summarized in Table~\ref{radioparam}.

Radio pulses were first searched for by dedispersing the GBT and Parkes telescope data both at the dispersion measure
of PSR~J1819$-$1458, 196.0~${\rm cm}^{-3}~{\rm pc}$, and with zero dispersion
using the SIGPROC\footnote{http://sigproc.sourceforge.net} pulsar processing package.
Zero-DM time series were created for the GBT and Parkes telescope data
to help discriminate pulses from terrestrial radio sources.
We could not dedisperse the Effelsberg telescope data because it had only one frequency channel.
The radiometer noise, which is the root-mean-square deviation of the time series
in flux density units, determines the sensitivity of each observation and is given by             
\begin{equation}   
  \sigma = \frac{\beta T_{\rm sys}}{G \sqrt{n_{\rm p} t_{\rm samp} \Delta f}},
  \label{radiometer}
\end{equation}
where $\beta$ is a correction factor accounting for the loss in sensitivity due to digitization
(1.25, 1.16, and 1.00 for one-, three-, and sixteen-bit digitization of the Parkes, GBT, and Effelsberg 
Telescopes, respectively),
$T_{\rm sys}$ is the system temperature
(we included scaled 408-MHz sky temperatures of \citet{hks+81} assuming a spectral index of $-$2.6 \citep{lmop87}
in the values quoted in Table~\ref{radioparam}),
$G$ is the gain,
$n_{\rm p}$ is the number of polarizations summed (two in our case),
$t_{\rm samp}$ is the sampling time,
and $\Delta{}f$ is the bandwidth of the observation.
The parameters for each observation are detailed in Table~\ref{radioparam}.
The effective sampling time of the Effelsberg telescope of 46~ms listed in Table~\ref{radioparam}
is the dispersion delay of the pulse over the single frequency channel's bandwidth of its receiver.
Since this effective sampling time makes the Effelsberg telescope's 
radiometer noise misleadingly lower, we also provide a modified radiometer noise for comparison, $\sigma_{\rm 1ms}$,
which uses $t_{\rm samp}=1$~ms.
Radio pulses which were detected with higher SNR at the DM of the source than at zero DM
(for the GBT and Parkes telescope data),
exceeded the radiometer noise by a factor of five considering the false-alarm statistics,
and were in phase with the radio ephemeris were considered real.

\begin{figure*}
  \includegraphics[angle=0,width=7.0in]{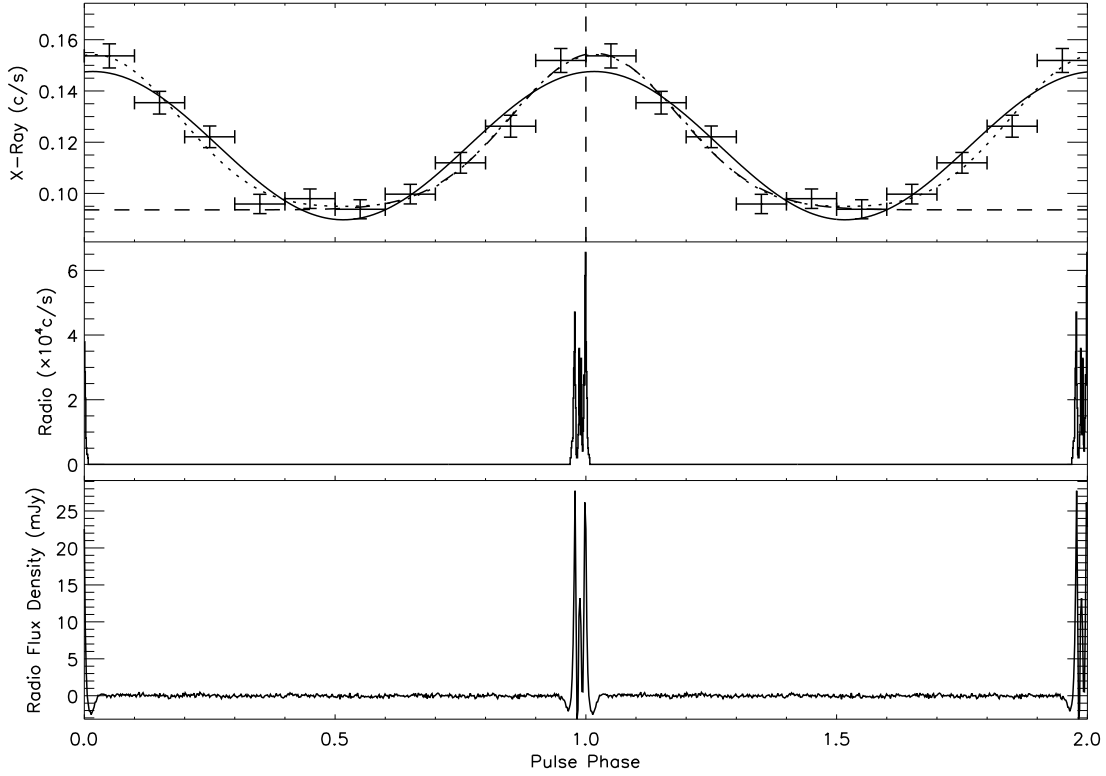}
  \caption{ X-ray and radio profiles of PSR~J1819$-$1458 folded using the radio ephemeris.
    {\it Top:} The background-corrected X-ray profile consists of ten phase bins over one rotational period,
    which consists of {\it XMM-Newton} PN and MOS detected photons within the 0.5~keV~$<~E~<$~2.6~keV energy range
    within the GTIs and PATTERN $\le12$, summing up to $\sim$17~hours of observation time.
    The horizontal and vertical bars indicate the size of the phase bins and the $\sqrt{N}$ errors.
    The solid, dotted, and dashed lines indicate the single sinusoid, two sinusoid,
    and Gaussian fits to the profile (fit over the 0.5$-$1.5 phase range), respectively.
    Note that the dotted and dashed lines overlap considerably.
    The vertical dashed line indicates the peak of the radio pulse profile (phase = 1.0).
    {\it Middle:} Radio pulse count histogram created by using the radio ephemeris to
    assign a phase to each barycentered pulse detected by the 7.7~hour observation of the GBT at an observing frequency of 2~GHz,
    and then binning all the radio pulse arrival times into a 2048 bin histogram.
    {\it Bottom:} Radio flux density profile formed from pulses detected using the 7.7~hour observation of the GBT
    at an observing frequency of 2~GHz.
    Flux densities were calculated by normalizing the scale of each detected pulse's
    off-pulse noise to the radiometer noise, then averaging all the pulses together.
    The dips preceding and following the pulse are due to digitization of the signal \citep[e.g.][]{ja98}.
    The profile is shown twice in all plots for clarity.
  }
  \label{fig_profile}
\end{figure*}

\begin{figure*}
  \includegraphics[angle=0,width=7.0in]{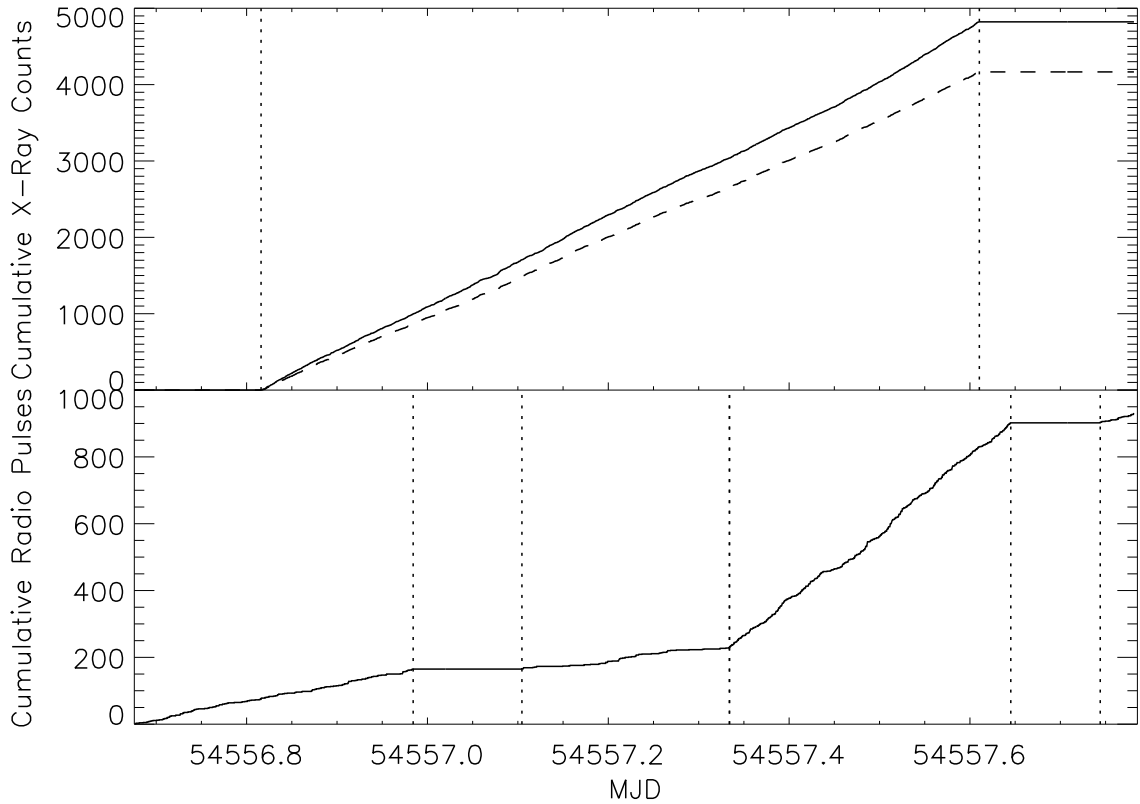}
  \caption{{\it Top:} Cumulative count of X-ray photon detections from PSR~J1819$-$1458.
    The solid line represents all X-ray photons while
    the dashed only includes the photons with energies within the range 0.5$-$2.6~keV.
    Dashed vertical lines designate the beginning and end of the GTIs.
    In both cases, we only include photons from the source region described in Section~\ref{timing}.
    {\it Bottom:} Cumulative radio pulses detected by the following radio telescopes over
    time - Parkes, Effelsberg, GBT, and then Parkes again.
    Dashed vertical lines indicate the beginnings and endings of the radio telescopes' observing time.
    Different radio observing frequencies and sensitivities bring about the different slopes of
    the cumulative radio pulse distribution.
    The two flat regions of the distribution are attributed to the times
    when the pulsar was not observed.
  }
  \label{cumulative_pulses}
\end{figure*}

The times of arrival of the pulses from the RRAT were converted to barycentered arrival times 
at infinite frequency using TEMPO and the X-ray derived position.
The folded solar system barycentered times are shown in the middle panel of Figure~\ref{fig_profile},
which was created by finding the phase of each pulse using the radio ephemeris and then
binning all the radio pulse arrival times into a 2048 bin histogram.
While individual radio pulses of PSR~J1819$-$1458 typically consists of a single
narrow pulse, the averaged radio pulse shape,
shown in the bottom panel of Figure~\ref{fig_profile},
has three separate components \citep{lmk+09,khsm09},
a center component of more fainter pulses and two outer components
made of fewer brighter pulses.
Each outer component is $\sim$45~ms apart from the center component,
much smaller than one of the ten bins in the top panel of Figure~\ref{fig_profile},
and does not affect the correlation analysis in Section~\ref{correlate}
since only correlations greater than one spin period are considered.
We detected 165 radio pulses in the first Parkes observation (i.e. 21 pulses/hour),
64 pulses in the Effelsberg observation (i.e. 12 pulses/hour),
673 pulses in the GBT observation (i.e. 90 pulses/hour),
and 29 pulses in the second Parkes observation (i.e. 29 pulses/hour) 
for 931 radio pulses (bottom panel of Figure~\ref{cumulative_pulses}).

\section{Correlation of Radio Pulses and X-ray Photons}
\label{correlate}

\begin{figure*}
  \includegraphics[angle=0,width=7.0in]{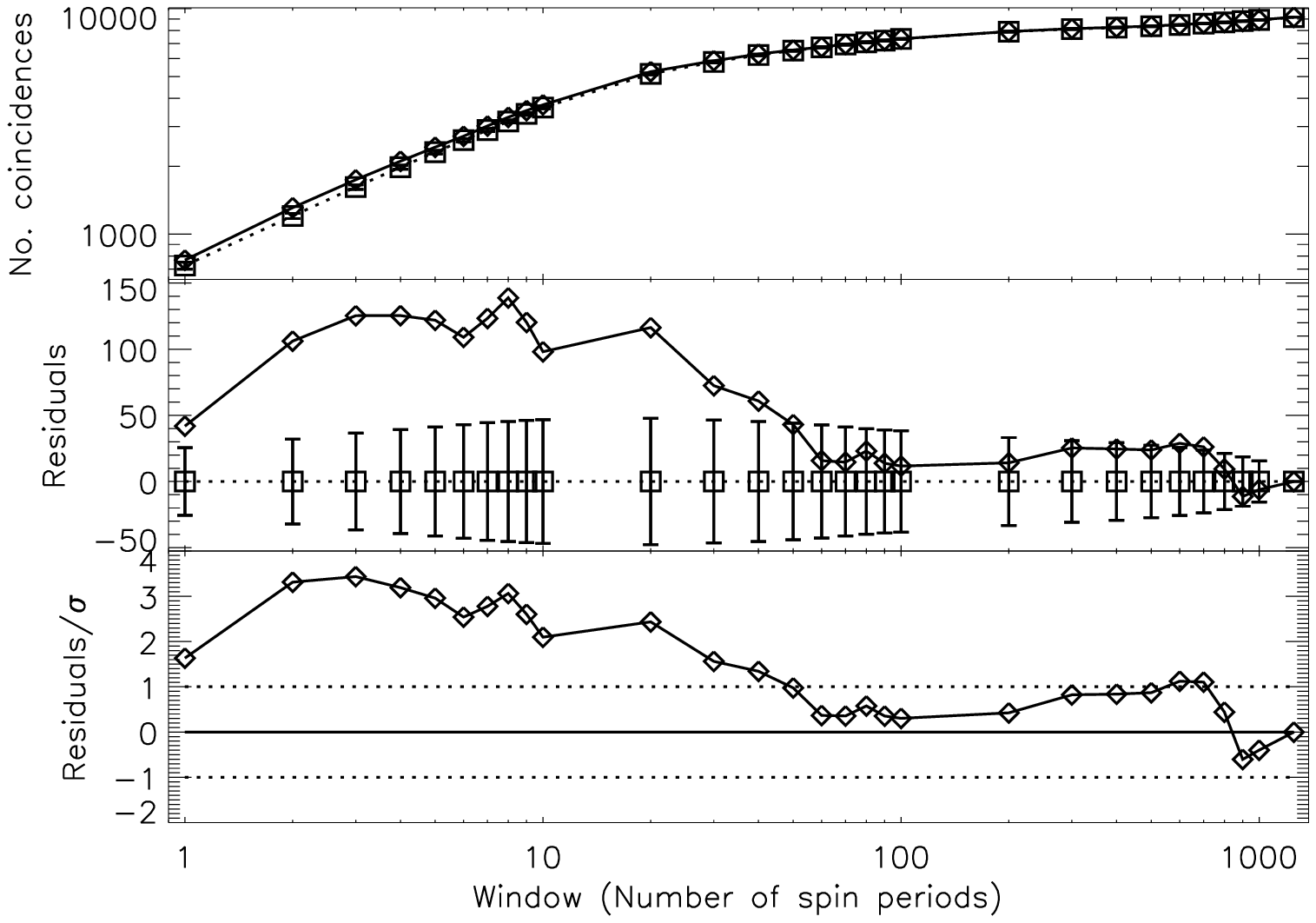}
  \caption{
    {\it Top:} Comparison of the number of PATTERN $\le12$ X-ray photons 
    from PSR~J1819$-$1458 within the 0.5$-$2.6~keV energy range of the PN and MOS cameras
    that are coincident with a radio pulse within a given search window.
    X-ray photons coincident with the radio pulses are represented by diamonds and a solid line
    while the results of our simulation are represented by the squares (mean coincident photons of the simulations),
    vertical bars (standard deviation of coincident photons of the simulations) and the dotted line.
    {\it Middle:} The difference in the number of coincident photons for each window size in the data and the mean
    of the simulations.
    {\it Bottom:} As the middle plot, normalized by the standard deviation of the simulations.
    Here the horizontal dotted line indicates one standard deviation of the simulated random sets.
  }
  \label{coincident}
\end{figure*}

For the correlation analysis we only considered X-ray photons from the 0.5$-$2.6~keV energy range determined in Section~\ref{timing},
shown as the dashed line in Figure~\ref{cumulative_pulses}.
Analysis of the X-ray events within the GTIs satisfying 
either PATTERN $=0$ (i.e. allowing for only single events) as well as PATTERN $\le12$
was performed to see if this had any effect on the result.
The PATTERN $=0$ requirement, not shown, 
yielded 4166 PN events, 1425 MOS1 events, and 1512 MOS2 events for a total of 7103 detections.
and PATTERN $\le12$ requirement, shown in Figure~\ref{coincident}, 
yielded 5692 PN events, 1705 MOS1 events, and 1767 MOS2 events for a total of 9164 detections.

The radio coverage described in Section~\ref{radio} was not continuous due to two gaps --
one between the first Parkes and Effelsberg observation
and one between the GBT and second Parkes observation.
The second Parkes telescope observation was contemporaneous with our {\it XMM-Newton} observation,
but that portion of the {\it XMM-Newton} data was completely excluded by the GTIs.
Due to the discontinuous radio and X-ray observation coverage as well as differences in radio telescope sensitivities,
the distribution of time delays between X-ray detections and radio pulse detections is non-Gaussian.
In order to measure the significance of any correlations between detected radio pulses and X-ray photons from PSR~J1819$-$1458,
we created a series of random X-ray photon distributions that would be consistent with the  discontinuous coverage.
We distributed the photon times throughout the GTIs, sampling from a flat (random) distribution.
We created an array of $10^4$ random X-ray distributions to then compare to the radio pulse arrival times, 
in addition to the comparison with the XMM-Newton data.

We calculated the number of X-ray events detected by the PN and MOS cameras
coincident with detected radio pulses at different lag times (see Figure~\ref{coincident}).
The lag time for each X-ray photon was calculated as
the time elapsed between the X-ray detection time and its nearest detected radio pulse,
either before or after the X-ray detection.
In this case, an X-ray detection was considered coincident if there was
a radio pulse detected within some specified window of time,
e.g. for a window of ten periods an X-ray photon was counted as 
coincident if there was at least one radio pulse detected within
$10\times4.26$~s $=$ 42.6~s of either before or after the X-ray event.
This was also done for the array of simulated random sets.
Then the mean and standard deviation of these sets for each lag window were calculated
(the squares with vertical error bars in Figure~\ref{coincident}).
Differences between the data and simulated random sets
are shown on the middle plot of Figure~\ref{coincident}.
Finally, the differences were divided by the standard deviations of the simulated random sets,
shown in the bottom plot of Figure~\ref{coincident}.
For most lag window sizes, the number of coincident X-ray photons in the data exceeds the 
number of coincident X-ray photons in the simulated random sets.
The largest deviation between the data and the simulations 
is $3.2\sigma$ and $3.4\sigma$ at $3P \approx 12.8$~s
(where $P$ is the rotational period of the pulsar)
for the PATTERN = 0 and PATTERN $\le12$ cases, respectively.
Specifically, there were 1352 coincident X-ray detections but a mean of only 1262 coincident photons
from the simulated random sets with a standard deviation of 32 photons with a $3P$ window size for the PATTERN = 0 case
and 1742 coincident X-ray detections but only a mean of 1617 coincident photons
from the simulated random sets with a standard deviation of 36 photons with a $3P$ window size for the PATTERN $\le12$ case.
Of our $10^4$ random sets, only 46 sets had a deviation exceeding $3.4\sigma$ for one or more window sizes, 
the probability of this occurring by chance is then 0.46\%.
Note that as the window size gets large enough,
the data and simulated data sets converge once all the photons are considered coincident.
 
\begin{figure*}
  \includegraphics[angle=0,width=7.0in]{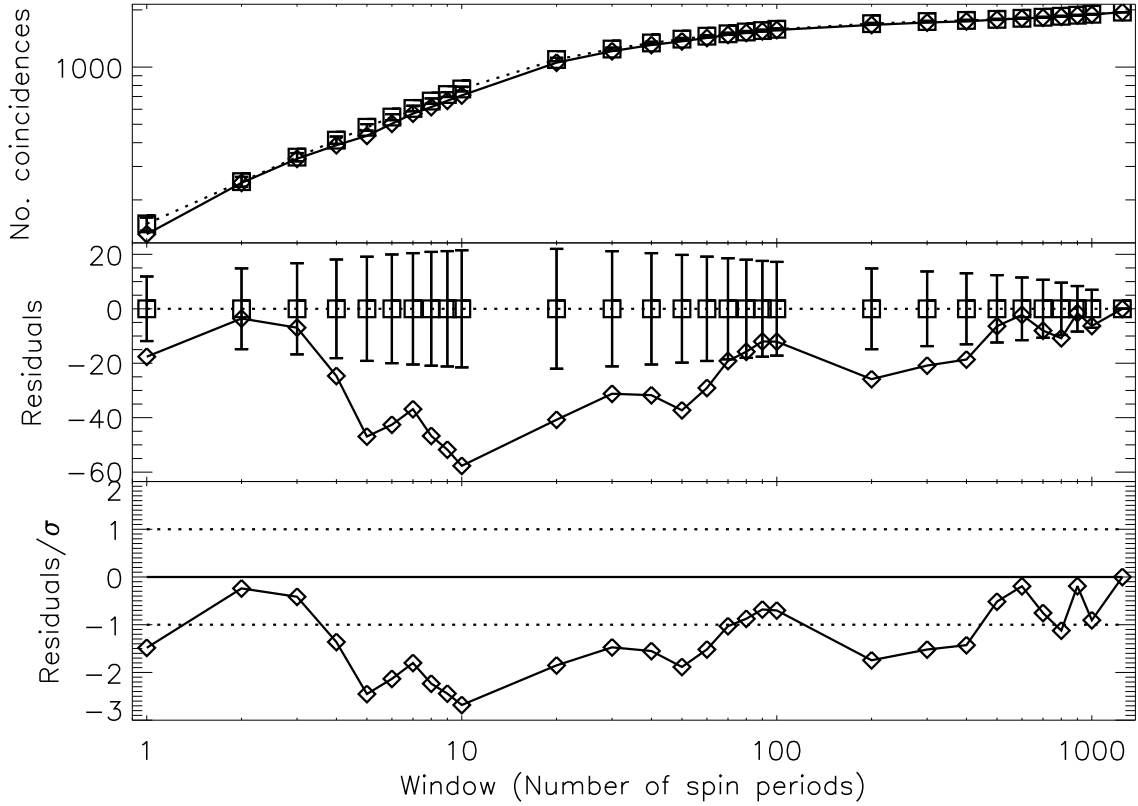}
  \caption{
    {\it Top:} Comparison of the number of X-ray photons from 2XMMi~J181928.8$-$145202 within the 0.5$-$2.6~keV energy range
    from the PN detector only
    that are coincident with a radio pulse within a given search window.
    X-ray photons coincident with the radio pulses are represented by diamonds and a solid line
    while the results of $10^4$ simulations are represented by the boxes (mean coincident photons of the simulations),
    vertical bars (standard deviation of coincident photons of the simulations) and the dotted line.
    {\it Middle:} The difference between the number of coincident photons for each window size in the data and the mean
    of the simulations.
    {\it Bottom:} As in the middle plot, normalized by the standard deviation of the simulations.
    Here the horizontal dotted line indicates one standard deviation of the simulated random sets.
  }
  \label{coincident_off}
\end{figure*}

To help us gauge the significance of these deviations of the data versus randomized times, 
we also did the same analysis for another source
on the same CCD in the field of view of {\it XMM-Newton},
2XMMi~J181928.8$-$145202\footnote{http://xmmssc-www.star.le.ac.uk/Catalogue/2XMMi/},
% a.k.a. ARXA~J181929.0--145201 a.k.a. X181928.88--145206.7 a.k.a. 2MASS~18192899--1452043
(see Figure~\ref{coincident_off}).
This source was not visible on the MOS1 detector, but we only considered the PN detector for this comparison.
In this case, the randomized times are coincident more often than the real data,
with the largest deviation peaking at $2.7\sigma$ below the mean of the simulations at $10P = 42.63$~s.
Of these $10^4$ random sets, 408 sets had a deviation exceeding $2.7\sigma$ for one or more window sizes,
the probability of this occurring by chance is then 4.08\%.

The Kolmogorov$-$Smirnov (KS) test (see, e.g., Press et~al. 1986 \nocite{pftv86}) 
was used to determine the degree to which the X-ray data set itself differs from a random distribution.
In this case we used both the numerical recipes {\tt ksone},
which compares a single data set to an analytical distribution,
and {\tt kstwo}, which compares two data sets to one another.
When comparing the combined PN and MOS detections to a flat distribution throughout the GTIs,
the KS statistic from {\tt ksone} is 0.14
(note that small values indicate the set is significantly different from the distribution).
When we compared our $10^4$ simulated random sets to the distribution with {\tt ksone},
we found a mean KS statistic of $0.5\pm0.3$.
We also compared the PN and MOS detections to the array of simulated random sets using {\tt kstwo}.
In this case, the mean KS statistic of these comparisons is $0.3\pm0.3$,
where the $\pm0.3$ represents the standard deviation of the $10^4$ cases.
We compared the radio pulse detections at each observation to a random distribution.
The KS statistic from {\tt ksone} was 0.77, 0.07, 0.09, and 0.23 for
the first Parkes, Effelsberg, GBT, and second Parkes observations, respectively.
We then compared each observation's pulse detections to $10^4$ simulated random sets
with flat distributions containing the same number of pulse detections using {\tt kstwo}.
The mean KS statistic of these comparisons is
$0.6\pm0.3$, $0.5\pm0.3$, $0.5\pm0.3$, and $0.5\pm0.3$
for the first Parkes, Effelsberg, GBT, and second Parkes observations, respectively.
These statistics show that individually the X-ray photons and the radio pulse
detections are consistent with random distributions.

\section{Conclusions}
\label{conclusions}

We have observed concurrent X-ray and radio pulsations from PSR~J1819$-$1458.
The peak of the X-ray profile is offset from the radio profile by $0.02\pm0.01$ in phase,
which means they occur at the same phase within the timing resolution of {\it XMM-Newton} 
(73.4~ms, or 0.017 of the period).
There is also evidence of a second sine-wave at twice the
rotational frequency of the radio pulses and aligned with the X-ray profile peak,
suggesting X-ray emission from the other pole of the neutron star
and befitting a two blackbody model with both poles as hotspots.
This is consistent with radio polarization observations
which show that PSR~J1819$-$1458 could be an orthogonal rotator
(the angle between the pulsar's rotational axis and its magnetic dipole axis, $\alpha$,
is not well-constrained, but most likely has a value near $90\degree$, see \citet{khsm09}).

The spectrum is consistent with a thermal emitter with
a broad absorption line, possibly composed of two different lines around $\sim$1.0 and $\sim$1.3~keV.
Coupled with the detection of the absorption seen in a previous {\it XMM-Newton} observation \citep{mrg+07}
and in the {\it Chandra} data \citep{roz+05,crb+12}, we are certain of its astrophysical nature.
If the line is due to proton resonant cyclotron scattering,
then the cyclotron absorption line at 0.907~keV 
({\tt BB$\times$Cyclotron} in Table~\ref{spec})
implies a dipole magnetic field strength of $1.9\times10^{14}$~G.
If the absorption line is due to electron resonant cyclotron scattering,
then the dipole magnetic field strength would be closer to $1\times10^{11}$~G.
The surface dipole magnetic field strength estimate is proportional to the cosecant of $\alpha$ \citep{lk05}.
The surface dipole magnetic field strength is then consistent with the
cyclotron proton resonant scattering model for $\alpha = 15\degree$.
As we have described above there is evidence that $\alpha$ may be closer to $90\degree$,
though $\alpha$ is not well constrained. 
The inferred $\alpha$ for the electron cyclotron case is undefined 
since the implied dipole magnetic field strength of the model would be weaker than the surface
dipole magnetic field strength estimate. 
This makes the electron cyclotron model unlikely.

We fit a blackbody temperature of $kT\sim$0.14~keV, 
slightly higher than what is expected from fast cooling models for
high magnetic field pulsars \citep{apm08,pmg09}.
This relation, however, assumes the spin-down age to be correct,
which might not be true \citep{nsk+13},
especially given the unusual glitch behavior of PSR~J1819$-$1458 \citep{lmk+09}.
It is also interesting to consider that the derived X-ray luminosity
from our best-fit model, a blackbody model with two Gaussian absorption lines,
is $L_{\rm 0.3-5.0 keV} \sim 3\times10^{33}$~ergs~s$^{-1}$,
which exceeds the pulsar's spin-down luminosity by a factor of $\sim10$.
The $>25$\% uncertainty in the distance estimate, however, 
lends an even larger uncertainty to the derived X-ray luminosity estimate.
The temperature and possibly high luminosity, combined with the unusual glitch activity,
suggests that it could be a transitional object between pulsars and magnetars.

Our KS test results show that both the X-ray photon
and radio pulse detections are consistent with random distributions.
However, we have shown that the X-ray photon and radio pulse detections
may be correlated on timescales of less than 10 pulsar spin periods,
where we measured a $3.4\sigma$ deviation in our data from random distributions.
As mentioned in Section~\ref{intro}, this tentative correlation suggests a link between the physical process producing the radio pulses 
and the heating of the polar-cap 
and represents the first enhancement of X-ray emission associated with radio pulse variability.

\citet{zgd07} proposed two interpretations which may explain the relationship between 
nulling pulsars, RRATs, and conventional radio pulsars.  
Their first model interpreted
RRATs and nulling pulsars as dead pulsars which sporadically re-activate
when coherent emission and pair production conditions are met.
Their second model interpreted RRATs'
behavior as a complement to nulling pulsars 
undergoing a reversal of radio emission direction.
Zhang et~al. proposed that X-ray observations
may help discern between the two interpretations
and specifically mention PSR~J1819$-$1458
as fitting within the re-activated dead pulsar model
because of its apparent lack of a non-thermal component in its X-ray spectrum \citep{rbg+06,gmr+07}.
Even though we are currently unable to
constrain a power-law tail, the tentative correlation between the
radio pulse and X-ray photon detection times suggests the reactivation
model for PSR~J1819$-$1458.

\acknowledgments
The National Radio Astronomy Observatory is a facility of the National Science Foundation operated under cooperative agreement by Associated Universities, Inc.
JM was supported by an NRAO student support grant.
JM and MAM were supported by NASA {\it XMM-Newton} observer support award NNX10AD14G.
MAM gratefully acknowledges support from Oxford Astrophysics while on sabbatical leave.
NR acknowledges support from a Ramon y Cajal fellowship and grants AYA2009$-$07391, AYA2012$-$39303, SGR2009$-$811, TW2010005 and iLINK2011$-$0303.

\bibliographystyle{apj}
\bibliography{journals,psrrefs,modrefs,crossrefs}

\newpage

%Figure 1

%Figure 2

% Figure 3

%Figure 4

%Figure 5

%%%%%%%%%%%%%%%%%%%%%%%%%%%%%%%%%%%%%%% TABLE 1 - GTIs %%%%%%%%%%%%%%%%%%
\begin{table*}
\begin{center}
\caption{X-ray Good Time Intervals}
%\vspace{0.3cm}

\noindent\makebox[\textwidth]{%
\begin{tabular}{lccc}
\hline
\hline
               & PN                            & MOS1                          & MOS2 \\
\hline
GTI~1 MJD span & 54556.8164062$-$54556.8496346 & 54556.8164062$-$54556.8494625 & 54556.8164062$-$54556.8503720 \\
GTI~2 MJD span & 54556.8496771$-$54556.8504032 & 54556.8496430$-$54556.8503653 & 54556.8505225$-$54557.6338842 \\
GTI~3 MJD span & 54556.8505407$-$54557.6093750 & 54556.8505158$-$54557.6093750 & 54557.6339443$-$54557.6093750 \\
\hline
\end{tabular}}
\label{GTItable}
\tablecomments{Good Time Intervals (GTIs) for the PN, MOS1, and MOS2 detectors. See Section~\ref{xrayprops} for details.}
\end{center}
\end{table*}

%%%%%%%%%%%%%%%%%%%%%%%%%%%%%%%%%%%%%%% TABLE 2 - X-ray Spectrum %%%%%%%%%%%%%%%%%%
\begin{table*}
\begin{center}
\caption{Spectral fits for PSR~J1819$-$1458 with EPIC-PN}
\vspace{0.3cm}

\resizebox{7.0in}{!} {

\begin{tabular}{lcccccc}
\hline
\hline
                  & Blackbody (BB)      & BB$\times$Neon        & BB$\times$Gaussian & BB$\times$Cyclotron & BB+BB & BB$\times$Gaussian$\times$Gaussian \\
\hline
$N_{\rm H}$       & 0.9$\pm0.1$         & 0.88$^{+0.10}_{-0.08}$ & 1.244$\pm0.009$   & 1.174$\pm0.009$ & 1.4$^{+0.3}_{-0.2}$ & 0.88$\pm0.01$      \\
$N_{O}$           & 0.3$^{+0.2}_{-0.3}$ & 0.7$^{+0.2}_{-0.3}$    & 0.82$\pm0.02$     & 0.87$\pm0.02$     & 0.8$^{+0.1}_{-0.2}$ & 0.69$\pm0.03$    \\
$N_{\rm Ne}$      & --                  & 3.0$\pm0.7$            & --                & --                & --                  & --                 \\
$E_{\rm cy}$      & --                  & --                     & --                & 0.907$\pm0.009$   & --                  & --                \\
$w_{\rm cy}$      & --                  & --                     & --                & 0.54$\pm0.02$     & --                  & --                 \\
$d_{\rm cy}$      & --                  & --                     & --                & 1.32$\pm0.02$     & --                  & --                 \\
$E_{\rm G1}$      & --                  & --                     & 1.12$\pm0.01$     & --                & --                  & 1.00$\pm0.01$   \\
$\sigma_{\rm G1}$ & --                  & --                     & 0.39$\pm0.01$     & --                & --                  & 0.004$\pm0.001$   \\
$\tau_{\rm G1}$   & --                  & --                     & 1.41$\pm0.03$     & --                & --                  & 4$^{+51}_{-3}$    \\
$E_{\rm G2}$      & --                  & --                     & --                & --                & --                  & 1.29$\pm0.03$    \\
$\sigma_{\rm G2}$ & --                  & --                     & --                & --                & --                  & 0.18$\pm0.03$    \\
$\tau_{\rm G2}$   & --                  & --                     & --                & --                & --                  & 0.18$\pm0.02$     \\
$kT_{1}$          & 0.140$\pm0.005$     & 0.131$\pm0.006$        & 0.1133$\pm0.0005$ & 0.1312$\pm0.0007$ & 0.07$\pm0.01$       & 0.1382$\pm0.0009$  \\
$kT_{2}$          & --                  & --                     & --                & --                & 0.15$^{+0.02}_{-0.01}$ & --             \\
Abs. Flux & 1.35$^{+0.02}_{-0.03}$      & 1.36$^{+0.02}_{-0.05}$ & 1.37$\pm0.03$     & 1.37$\pm0.03$     & 1.37$\pm0.03$       & 1.37$^{+0.07}_{-0.04}$  \\
Unab. Flux$_{1}$  & 13.3$\pm0.3$        & 25.5$\pm0.5$           & 224$\pm5$         & 155$\pm3$         & 520$\pm20$          & 21.8$\pm0.5$       \\
Unab. Flux$_{2}$  & --                  & --                     & --                & --                & 14.5$\pm0.7$        & --            \\
$R_{1}$           & 6$\pm4$             & 10$\pm6$               & 40$\pm20$         & 30$^{+10}_{-20}$  & 100$\pm100$         & 8$^{+5}_{-4}$    \\
$R_{2}$           & --                  & --                     & --                & --                & 6$^{+4}_{-3}$       & --          \\
$\chi^2_\nu$ (d.o.f.) & 1.41 (153)      & 1.28 (152)             & 1.21 (150)        & 1.19 (150)        & 1.19 (151)          & 1.09 (147)      \\
\hline
\hline
\end{tabular}
}
\label{spec}
\tablecomments{
  Parameters fit to our data combined with the \citet{mrg+07} data,
  fitting in the 0.5$-$2.0~keV energy range.
  Fluxes are calculated in the 0.3$-$5.0~keV energy range
  for direct comparison to the observation done by \citet{mrg+07},
  reported in units of $10^{-13}$~ergs~s$^{-1}$~cm$^{-2}$.
  $N_{\rm H}$ is in units of $10^{22}$cm$^{-2}$
  while $N_{\rm O}$ and $N_{\rm Ne}$ are in solar units (assuming solar abundance from \citet{l03}).
  The photoelectric cross-section of Verner et~al. (1998) has been used for all fits.
  The values of $kT$ (blackbody temperature), $E_{\rm G}$ (Gaussian line energy),
  $\sigma_{\rm G}$ (Gaussian line width),
  $E_{\rm cy}$ (cyclotron line energy) and $w_{\rm cy}$ (cyclotron line width)
  are in units of keV.
  $R_1$ (blackbody emission radius at infinity assuming a 3.6~kpc distance) 
  and $R_2$ (blackbody hotspot emission radius at infinity in the two blackbody model, also assuming a 3.6~kpc distance)
  are in units of km.
  The Gaussian line depth $\tau_{\rm G}$
  and fundamental cyclotron line depth $d_{\rm cy}$ are dimensionless.
  Errors are at the 1$\sigma$ confidence level.
  XSPEC models used are (from left to right):
  {\tt vphabs*bbody}, {\tt vphabs*bbody}, {\tt vphabs*gabs*bbody}, {\tt vphabs*cyclabs*bbody},
  {\tt vphabs*(bbody+bbody)}, and {\tt vphabs*gabs*gabs*bbody}.
}
\end{center}
\end{table*}

%%%%%%%%%%%%%%%%%%%%%%%%%%%%%%%%%%%%%%% TABLE 3 - Radio Data %%%%%%%%%%%%%%%%%%
\begin{table*}
\begin{center}
\caption{Radio Parameters}
%\vspace{0.3cm}

\noindent\makebox[\textwidth]{%
\begin{tabular}{lcccc}
\hline
\hline
                          & Parkes 1 & Effelsberg & GBT        & Parkes 2 \\
\hline
MJD span                  & 54556.67$-$54557.00 & 54557.10$-$54557.33 & 54557.33$-$54557.65 & 54557.74$-$54557.78 \\
Center Freq. (GHz)        & 1.4      & 1.4        & 1.9        & 1.4 \\
Bandwidth (MHz)           & 256      & 80         & 600        & 256 \\
No. of frequency channels & 512      & 1          & 768        & 512 \\
Sampling Time ($\mu$s)    & 100      & 46000      & 81.92      & 100 \\
Observation Length (hr)   & 7.9      & 5.5        & 7.7        & 1.0 \\
$\beta$                   & 1.25     & 1.00       & 1.16       & 1.25 \\
$G$ (K/Jy)                & 0.67     & 1.5        & 1.9        & 0.67 \\
$T_{\rm sys}$ (K)         & 39       & 27         & 29         & 39 \\
$\sigma$ (mJy)            & 320      & 6.6        & 56         & 320 \\
$\sigma_{\rm 1ms}$ (mJy)  & 100      & 45         & 16.2       & 100 \\
\hline
\end{tabular}}
\label{radioparam}
\tablecomments{Radio observation parameters.  See Section~\ref{radio} for details.}
\end{center}
\end{table*}

\end{document}